\documentclass[journal=jacsat,manuscript=article,layout=twocolumn]{achemso}

\usepackage[version=3]{mhchem} 
\usepackage{amsmath}
\usepackage{amsfonts}
\usepackage[english]{babel}
\usepackage[T1]{fontenc}
\usepackage{graphicx}
\usepackage{epstopdf}
\usepackage{caption}
\usepackage{subcaption}
\usepackage{color}
\usepackage{array}
\usepackage{natbib}
\usepackage{mciteplus}
\usepackage{graphicx,dblfloatfix}
\usepackage{changes}
\usepackage{amssymb,bm}
\usepackage{abstract}
\setkeys{acs}{usetitle = true}
\author{S. Yu}
\affiliation{Soft Matter and Functional Materials, Helmholtz-Zentrum Berlin, Hahn-Meitner Platz 1, 14109 Berlin, Germany, and Helmholtz Virtual Institute "Multifunctional Biomaterials for Medicine", Kantstr. 55, 14513 Teltow, Germany }
\alsoaffiliation{Institut f{\"u}r Physik, Humboldt-Universit{\"at} zu Berlin, Newtonstr. 15, 12489 Berlin, Germany}
\author{X. Xu}
\affiliation{Soft Matter and Functional Materials, Helmholtz-Zentrum Berlin, Hahn-Meitner Platz 1, 14109 Berlin, Germany}
\author{C. Yigit}
\affiliation{Soft Matter and Functional Materials, Helmholtz-Zentrum Berlin, Hahn-Meitner Platz 1, 14109 Berlin, Germany}
\alsoaffiliation{Institut f{\"u}r Physik, Humboldt-Universit{\"at} zu Berlin, Newtonstr. 15, 12489 Berlin, Germany}
\author{M.~van der Giet}
\affiliation[Charit\'e - Universit\"atsmedizin Berlin]{Medizinische Klinik f\"ur Nephrologie, Charit\'e , Hindenburgdamm 30, Charit\'e -Campus Benjamin Franklin, 12203 Berlin}
\author{W.~Zidek}
\affiliation[Charit\'e - Universit\"atsmedizin Berlin]{Medizinische Klinik f\"ur Nephrologie, Charit\'e , Hindenburgdamm 30, Charit\'e -Campus Benjamin Franklin, 12203 Berlin}
\author{J.~Jankowski}
\affiliation[RWTH Aachen]{Institute for Molecular Cardiovascular Research IMCAR, RWTH Aachen, Pauvelsstr. 30, 52074 Aachen}
\author{J.~Dzubiella}
\affiliation{Soft Matter and Functional Materials, Helmholtz-Zentrum Berlin, Hahn-Meitner Platz 1, 14109 Berlin, Germany, and Helmholtz Virtual Institute "Multifunctional Biomaterials for Medicine", Kantstr. 55, 14513 Teltow, Germany}
\alsoaffiliation{Institut f{\"u}r Physik, Humboldt-Universit{\"at} zu Berlin, Newtonstr. 15, 12489 Berlin, Germany}
\author{M.~Ballauff}
\affiliation{Soft Matter and Functional Materials, Helmholtz-Zentrum Berlin, Hahn-Meitner Platz 1, 14109 Berlin, Germany, and Helmholtz Virtual Institute "Multifunctional Biomaterials for Medicine", Kantstr. 55, 14513 Teltow, Germany}
\alsoaffiliation{Institut f{\"u}r Physik, Humboldt-Universit{\"at} zu Berlin, Newtonstr. 15, 12489 Berlin, Germany}
\email{matthias.ballauff@helmholtz-berlin.de}
\keywords{Human Serum Albumin, Polyelectrolyte, ITC, etc.}
\title[demonstration]{Interaction of Human Serum Albumin with short Polyelectrolytes: A study by Calorimetry and Computer Simulation}
\begin{document}
\begin{tocentry}
\includegraphics{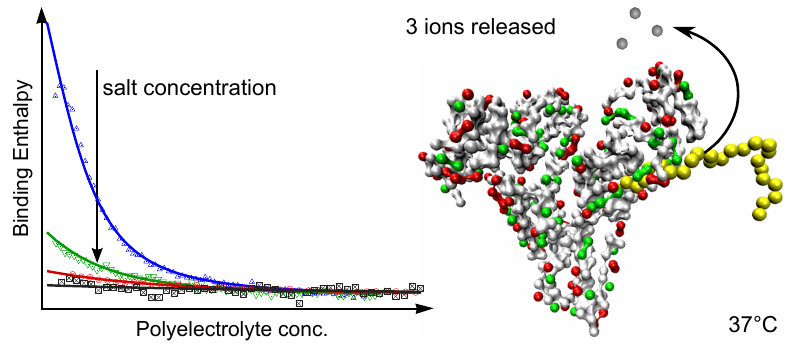}
\end{tocentry}
\captionsetup{font={sf,small}}
\twocolumn[
\maketitle 
\begin{onecolabstract}
We present a comprehensive study of the interaction of human serum albumin (HSA) with poly(acrylic acid) (PAA; number average degree of polymerization: 25) in aqueous solution. The interaction of HSA with PAA is studied in dilute solution as the function of the concentration of added salt (20 - 100~mM) and temperature (25 - 37$^{\circ}$C). Isothermal titration calorimetry (ITC) is used to analyze the interaction and to determine the binding constant and related thermodynamic data. It is found that only one PAA chain is bound per HSA molecule. The free energy of binding $\Delta G_b$ increases with temperature significantly. $\Delta G_b$ decreases with increasing salt concentration and is dominated by entropic contributions due to the release of bound counterions. Coarse-grained Langevin computer simulations treating the counterions in an explicit manner are used study the process of binding in detail. These simulations demonstrate that the PAA chains are bound in the Sudlow II site of the HSA. Moreover, $\Delta G_b$ is calculated from the simulations and found to be in very good agreement with the measured data. The simulations demonstrate clearly that the driving force of binding is the release of counterions in full agreement with the ITC-data.
\end{onecolabstract}
 ]

\section{Introduction}
Human Serum Albumin (HSA) is an important transport protein that interacts with substrates as e.g. fatty acids~\cite{Curry1998, Fasano2005} and pharmaceuticals in a specific manner.~\cite{Barbosa2005} Also, transport by HSA plays an important role for renal clearance and the interaction of HSA with typical uremic toxins \cite{Jankowski2003} as e.g. phenylacetic acid, indoxyl sulfate, and p-cresyl sulfate needs to be understood in detail.~\cite{Vanholder2003} Uremic toxins play an essential role in the excessive cardiovascular mortality and all cause mortality in patients with end stage renal disease. Elimination of these substances is essential to reduce the high cardiovascular disease risk.~\cite{Duranton2012,Duranton2013}
The obvious biological importance of this transport has led to a great number of studies of the interaction of HSA with various substrates, most notably by crystallography,~\cite{Bhattacharya2000, Matsuda2014} electron paramagnetic resonance spectroscopy, \cite{Reichenwallner2013} and isothermal titration calorimetry (ITC).~\cite{Chatterjee2012, Rehman2014} Thus, it is well established that the interaction of various substrates with HSA takes place mainly on the Sudlow I and the Sudlow II site.~\cite{Sudlow1975} ITC has been used to elucidate the thermodynamics of this interaction and by this the number of substrates adsorbed onto the surface of the protein.~\cite{Roselin2010,Bouchemal2008,Chatterjee2012,Keswani2010,Sivertsen2014}

Charge-charge interaction plays an essential role in the process of adsorption~\cite{Cooper2005, DaSilva2009, Becker2012} and a number of ITC-studies explore the dependence of the adsorption constant on ionic strength.~\cite{Ball2009,Seyrek2003,Du2014,Welsch2013} Recently, Jankowski and coworkers~\cite{Boehringer2015,Devine2014} have demonstrated that raising the ionic strength in the infusion fluid leads to an improved clearance of protein-bound toxins (PBT). Thus, raising the concentration of NaCl to 600~mM led to a significantly better removal of of uremic toxins as e.g. phenylacetic acid. This result points clearly to the central importance of Coulombic interactions for the binding strength of such toxins to HSA and to proteins in general. Charged toxins as phenylacetic acid are difficult to remove by conventional dialysis and exploring their interaction is of central importance for devising improved techniques for dialysis.~\cite{Boehringer2015}

Here we analyze the role of the interaction of charged substrates with HSA as the function of ionic strength and temperature. As a substrate we chose a short polyelectrolyte, namely poly(acrylic acid) (PAA) with 25 repeating units only. This substrate allows us to explore the Coulombic interaction of HSA with charged molecules in general. In addition to this, synergetic effects of adjacent carboxyl groups onto the binding constant can be elucidated. At the same time, PAA can be regarded as a model of charged toxins with molecular weight above 500~g/Mol.~\cite{Vanholder2003} ITC is used to obtain the binding constant and the number of bound PAA-molecules per protein and the temperature is varied between 25$^{\circ}$C and 37$^{\circ}$C. A first study of the interaction of polyelectrolytes with proteins by ITC has been presented by Schaaf~\textit{et al.} who demonstrated the general suitability of this method for the study of protein-polyelectrolyte interaction.~\cite{Ball2002}

The interaction of long polyelectrolytes with proteins in aqueous solution has been the subject of longstanding research that has led to an enormous literature.~\cite{Kayitmazer2013, Cooper2005, Nfor2008} Proteins can form complex coacervates with polyelectrolyte of opposite charge in aqueous solution and the strength of interaction is mediated by the ionic strength in the system.~\cite{Silva2010} If the ionic strength of the system is low enough, interaction may take place even on the ``wrong side`` of the isoelectric point, that is, proteins associate with polyelectrolytes of like charge. In many cases the formation of complexes between the protein and the polyelectrolyte is followed by precipitation and phase separation that may also lead to non-equilibrium states.~\cite{Du2014,Kayitmazer2013} Here we use a very short polyelectrolyte and low concentrations to avoid multiple interactions and phase separation. The results from the present experiments, however, may be directly utilized for a better understanding of complex coacervates.

Evidently, protein crystallography cannot be used to elucidate the structure of the complex between HSA and PAA. First of all, even in the bound state the non-bound segments of the polyelectrolyte will explore their conformational freedom in solution and the entropic contribution deriving therefrom will be an important part of the resulting change of free energy. Moreover, crystals of HSA will certainly not accommodate substrates of the size of a polyelectrolyte having 25 units. In order to make progress on a detailed structural investigation, we employ coarse-grained Langevin Dynamics (LD) computer simulations with an implicit solvent whereas the co- and counterions are treated in an explicit manner.~\cite{Clementi2000,Takada2012,Ravikumar2012} The protein is treated within the C$_\alpha$ - G$\bar{o}$ model, that is, each amino acid of HSA is modelled by a single bead bearing a charge or not. The polyelectrolyte PAA is also treated as a coarse-grained charged polymer. The combination of these models allows us to elucidate the details of the complex of PAA and HSA in solution with full structural resolution. In addition to this, LD-simulations allow us to calculate the free enthalpy of binding $\Delta G_b$ and to compare these data with measured values. Thus, the combination of ITC with computer simulation can be used to elucidate structural and thermodynamic details that are available by no other method.


\section{Experimental}\label{Sec:Experimental}
\textbf{Materials.} Polyacrylic acid (PAA) with M$_W$=1800~g/mol was purchased from Sigma-Aldrich (Schnelldorf, Germany) and used after several weeks of dialysis to match pH without changing ionic strength in the system. Human Serum Albumin (HSA) was also purchased from Sigma-Aldrich (lyophilized powder, Fatty acid free, Globulin free, $99\%$) with molecular weight calculated M$_W$=66400~g/mol and its purity verified by sds-gelelectrophoresis. The buffer morpholin-N-oxide (MOPS) was purchased from Sigma-Aldrich and used as received.

\subsection{Isothermal Titration Calorimetry} 
ITC experiments were performed using a VP-ITC instrument (Microcal, Northampton, MA). All samples were prepared in a pH 7.2 buffer solution using 10~mM MOPS and 10~mM, 40~mM, 60~mM and 90~mM NaCl to adjust ionic strength. All samples were dialyzed against buffer solution with according pH and degassed prior to experiment. For dialysis, the dialysis-system Float-a-Lyzer by Spectrum Labs with molecular weight cut-off (MWCO) 500-1000~Da for PAA and MWCO 20~kDa for HSA were used respectively. The samples were thermostatted and the instrument stabilized for 1 h to ensure thermal equilibrium and stability of the system. A total of 298~$\mu$L PAA solution was titrated with 75 successive 4~$\mu$L injections, stirring at 307~rpm and a time interval of 300-350~s between each injection into the cell containing 1.4~mL protein solution. The concentration of PAA and HSA were 1.3~g/L and 0.9~g/L respectively. We choose these low concentrations in order to study the interaction of single PAA-chains with only one HSA-molecule. Moreover, possible complications by coacervate formation are circumvented in this way. 
The experiments were performed at 25, 27.5, 30, 33, 37~$^o$C and ionic strengths 20~mM, 50~mM, 70~mM and 100~mM.

 As a first step of the ITC data analysis, the integration of the measured heat $Q$ over time is carried out to obtain the incremental heat $\Delta Q$ as a function of the molar ratio $x$ between polyelectrolyte and protein. For each experiment, the heat of dilution of PAA were measured separately by titrating the according polyelectrolyte into blank buffer solution and subtracted from adsorption heats. After correction, the resulting binding isotherm are fitted using a supplied module for Origin 7.0 (Microcal).\\
\subsubsection{Data analysis.}
\label{sec:itc_analysis}
 For the analysis, we chose the single set of independent binding sites (SSIS) model to fit all data.~\cite{Indyk1998} This model is based on the Langmuir equation, which assumes an equilibrium between the empty adsorption sites, the no. of proteins in solution and the occupied adsorption sites. 
This leads to the binding constant $K_b$: 
\begin{equation}
K_b=\frac{\Theta}{(1-\Theta)c_{PAA}}
\end{equation}
where $\Theta$ denotes the fraction of sites occupied by the polyelectrolyte and $c_{PAA}$ the concentration of free polyelectrolyte. Since only the total concentration of PAA $c_{PAA}^{tot}$ in the solution is known, $c_{PAA}$ is connected to the $c_{PAA}^{tot}$ as follows:
\begin{equation}
c_{PAA}^{tot}=c_{PAA}+N\Theta c_{prot}
\end{equation}
where $N$ is the number of free binding sites and $c_{prot}$ the total protein concentration in solution. Following these equations, the binding number $N$, binding affinity $K_b$ and the overall enthalpy change measured $\Delta H_{ITC}$ can be obtained by fitting the isotherm. In the following, we will show that only one PAA is adsorbed onto HSA in all cases. Therefore the parameter $N$ was fixed to unity for all subsequent fits. In this case the interaction of PAA and HSA can be formulated as a conventional chemical equilibrium:\\
\begin{equation}
K_b\approx\frac{c_{PAA-HSA}}{c_{PAA} (c^{tot}_{HSA}-c_{PAA-HSA}})
\label{eq:KbN1}
\end{equation}
where $c^{tot}_{HSA}$ denotes the total protein concentration in the solution and $c_{PAA-HSA}$ the concentration of PAA-HSA complex. Furthermore, the binding free energy $\Delta G_{b}$ can be derived by 
\begin{equation}
\Delta G_b=-RT\cdot ln{K_b}
\label{eq:DeltaG} 
\end{equation}
Furthermore using the following two equations, the entropy $\Delta S_b$ can be either calculated for one temperature or extracted from its temperature dependence:
\begin{equation}
\Delta G_b = \Delta H_{ITC}-T\Delta S
\label{eq:Gb_1}
\end{equation}
\begin{equation}
\frac{\partial\Delta G_b}{\partial T}=-\Delta S_b
\label{eq:Sb_vanthoff}
\end{equation}

\subsection{Theoretical Methods}
\subsubsection{Computer simulation model and parameters}
In our simulations we employ an implicit-water coarse-grained (CG) model, where each of the amino acids, PAA monomers, and salt ions is explicitly represented by a single interaction bead. Hence, the water is modeled by a dielectric background continuum while the salt is explicitly resolved.  The dynamics of each of the beads in our simulations is governed by Langevin's equation of motion~\cite{Ermak1978}
\begin{equation}
m_i\frac{d^2\bm{r}_i}{dt^2} = -m_i\xi_i\frac{d\bm{r}_i}{dt} + \bm\nabla_{i}U + \bm{R}_i(t)
\end{equation}
where $m_i$ and $\xi_i$ are the mass and friction constant of the $i$th bead, respectively. $U$ is
the system potential energy and includes harmonic angular and bonded interactions between
neighbouring beads in HSA and PAA, dihedral potential in the case of the HSA, and interatomic 
Lennard-Jones (LJ) between all non-neighbouring beads, including ions. Coulomb interactions govern the electrostatic pair potential between all charged beads. The random force $\bm{R}_i(t)$ is a
Gaussian noise process and satisfies the fluctuation-dissipation theorem
\begin{equation}
 \langle \bm{R}_i(t) \cdot \bm{R}_j(t') \rangle = 2m_i\xi_ik_BT\delta(t-t')\delta_{ij}.
\end{equation}
The simulations are performed using the GROMACS 4.5.4 software package.~\cite{Hess2008} A leap-frog
algorithm with a time step of 2 fs is used to integrate the equations of motion. A cubic box with side
lengths of $L = 30$ nm is employed and periodically replicated to generate a quasi-infinite system
in the canonical ensemble. The Langevin friction is {$\xi_i = 1.0$ ps$^{-1}$} that dissipates the energy at constant temperature $T$. 
Center of mass translation of the system is removed every 10 steps. The cutoff radius is set to {3.0 nm} to calculate the real-space interactions while Particle-Mesh-Ewald~\cite{Essmann1995} (PME) is implemented to account for long-range electrostatics. The PME method is computed in the reciprocal space with a FFT grid of {$\sim$~0.12 nm} spacing and a cubic interpolation of fourth-order.
The solvent is modelled as a continuous medium with a static dielectric constant of $\epsilon_r = 73.4$ and 78.2 for temperatures $T=25^{\circ}$C and 37$^{\circ}$C, respectively. All beads have mass $m_i = 1$~amu, diameter $\sigma_{LJ} = 0.4$~nm, energy well $\epsilon_{LJ} = 0.1$~k$_\text{B}$T and integer charges $q$ = 0, +1 or -1 e. The mass was chosen artificially low to enhance orientational fluctuations and sampling. Clearly, equilibrium properties, as investigated in this work, are not affected by any reasonable mass choices as long as the simulations are ergodic. 

The protein sequence for the HSA is provided by PDB databank: ID=1N5U.~\cite{Wardell2002} Every amino acid is described by a single bead positioned at its $C_{\alpha}$ atom. The native structure of the protein is maintained by a Go-model like force field as provided from the SMOG webtool for biomolecular simulations.~\cite{smog, calpha} All beads corresponding to basic and acidic amino acids are assigned a charge of +1$e$ and -1$e$, according to their titration state at physiological pH = 7.4, that is, ARG and LYS residues are assigned a charge +1$e$, while ASP and GLU are -1$e$, and HIS is neutral. 
With that the net charge of the simulated HSA is -14$e$. 

A single flexible PAA polyelectrolyte is modeled in a coarse-grained fashion as a sequence of $N_\text{mon}$ freely jointed beads. Each bead represents a monomer with a radius $\sigma_{LJ}$ and carries an electric charge of $-1e$. The PAA monomers are connected by a harmonic bond potential with an equilibrium bond length $b_\text{mon} = 0.4$~nm and a force constant $k_\text{mon} = 4100$ kJ mol$^{-1}$ nm$^{-2}$. The flexibility of the PAA chain is defined via a harmonic angle potential in which the angle between a triplet of monomer is $\phi = 120^\circ$ and the force constant is {$k_\phi = 418.4$} kJ mol$^{-1}$ rad$^{-2}$.
As in the experiments we consider short PAA chains with $N_\text{mon} = 25$ monomers. 
 The simulation box contains one HSA, while the amount of PAA and ions are characterized by the molar ratio $x = c_{PAA}/c_{HSA}$, ranging from 1 to 10, and salt concentration $c_{salt}$, ranging from 20 to 100~mM, respectively.     

\subsubsection{Binding and free energy calculations}
The stoichiometry, in our case the average number of bound PAA chains on one HSA, can be determined through a calculation of the normalized density distribution function $g(r)= c(r)/c^{tot}_{PAA}$, where $r$ is 
the distance  between the centers-of-mass (COM) coordinates of the HSA and PAA and $c^{tot}_{PAA}$ is the PAA bulk concentration. 
Integration of the $g(r)$  further leads to the PAA coordination number 
\begin{equation}
n(r) = 4\pi c^{tot}_{PAA}\int^{r}_{0} g(r') r'^2 dr'. 
\end{equation} 
which describes how many PAA chains are bound on average at a distance $r$. For each molar ratio we simulated 120 ns in order to calculate the equilibrium coordination (binding) numbers of PAA to HSA.   
 \begin{figure*}[!b]
\center
\includegraphics[width=0.3\textwidth]{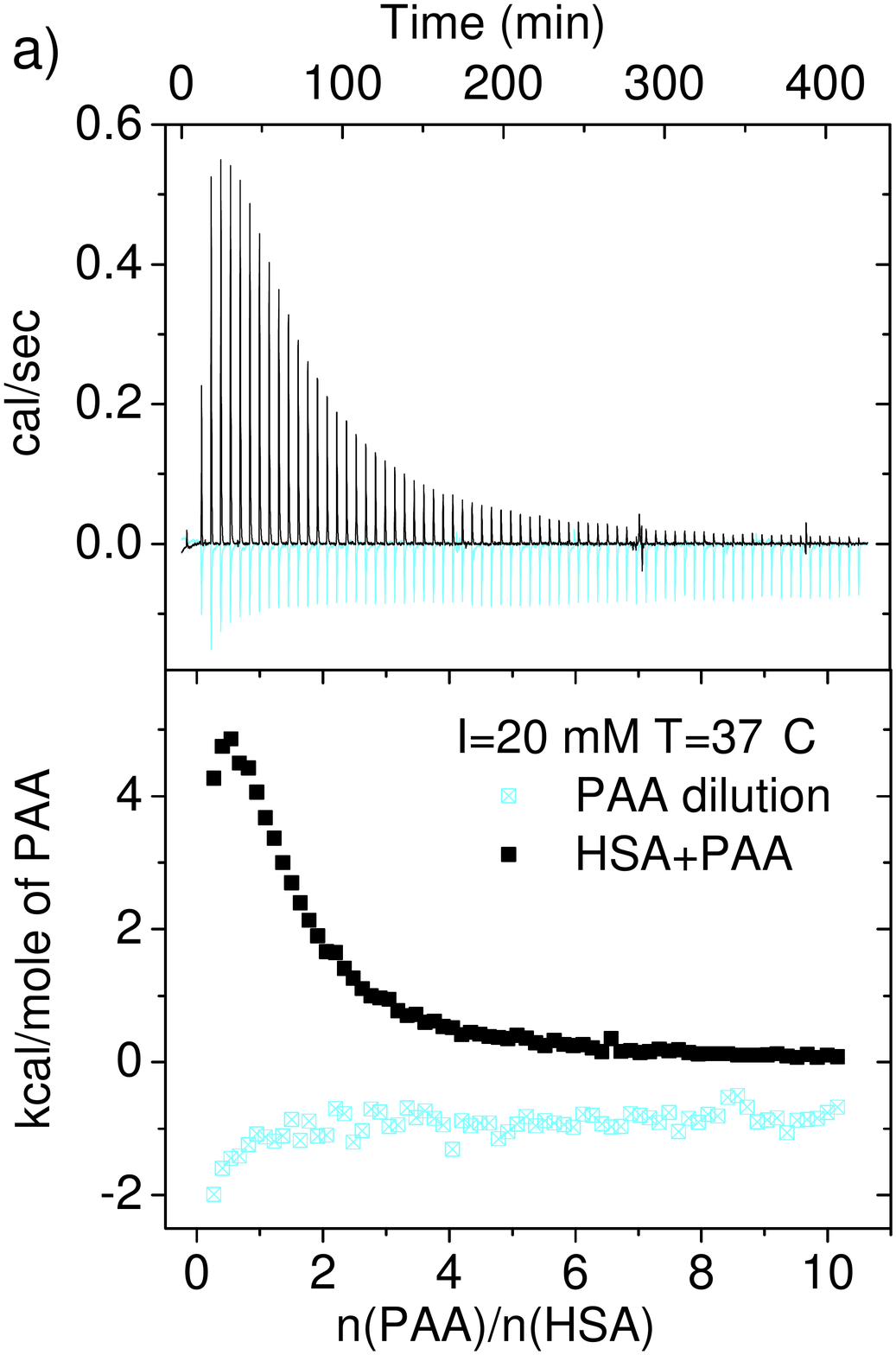}
\includegraphics[width=0.3\textwidth]{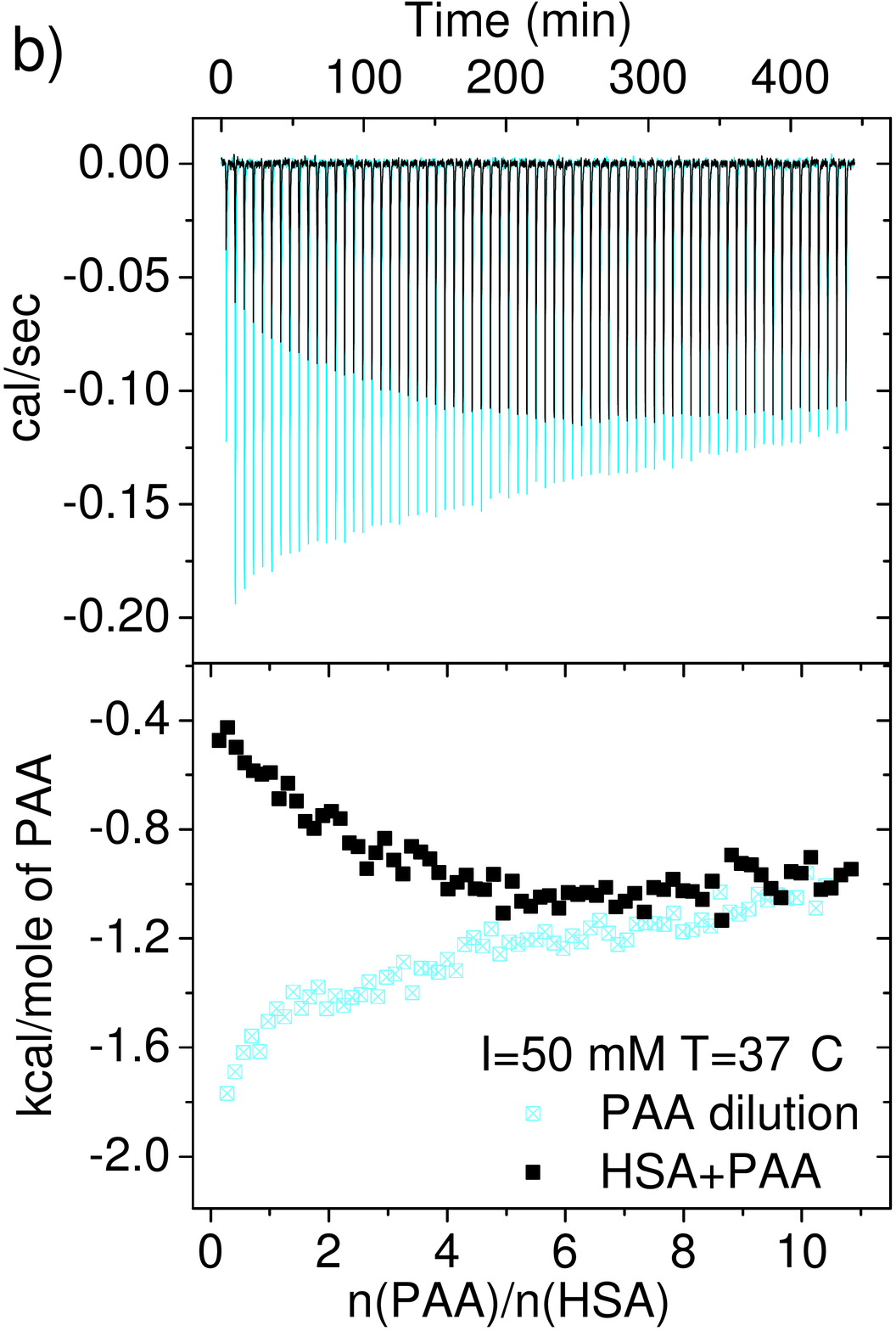}
\includegraphics[width=0.29\textwidth]{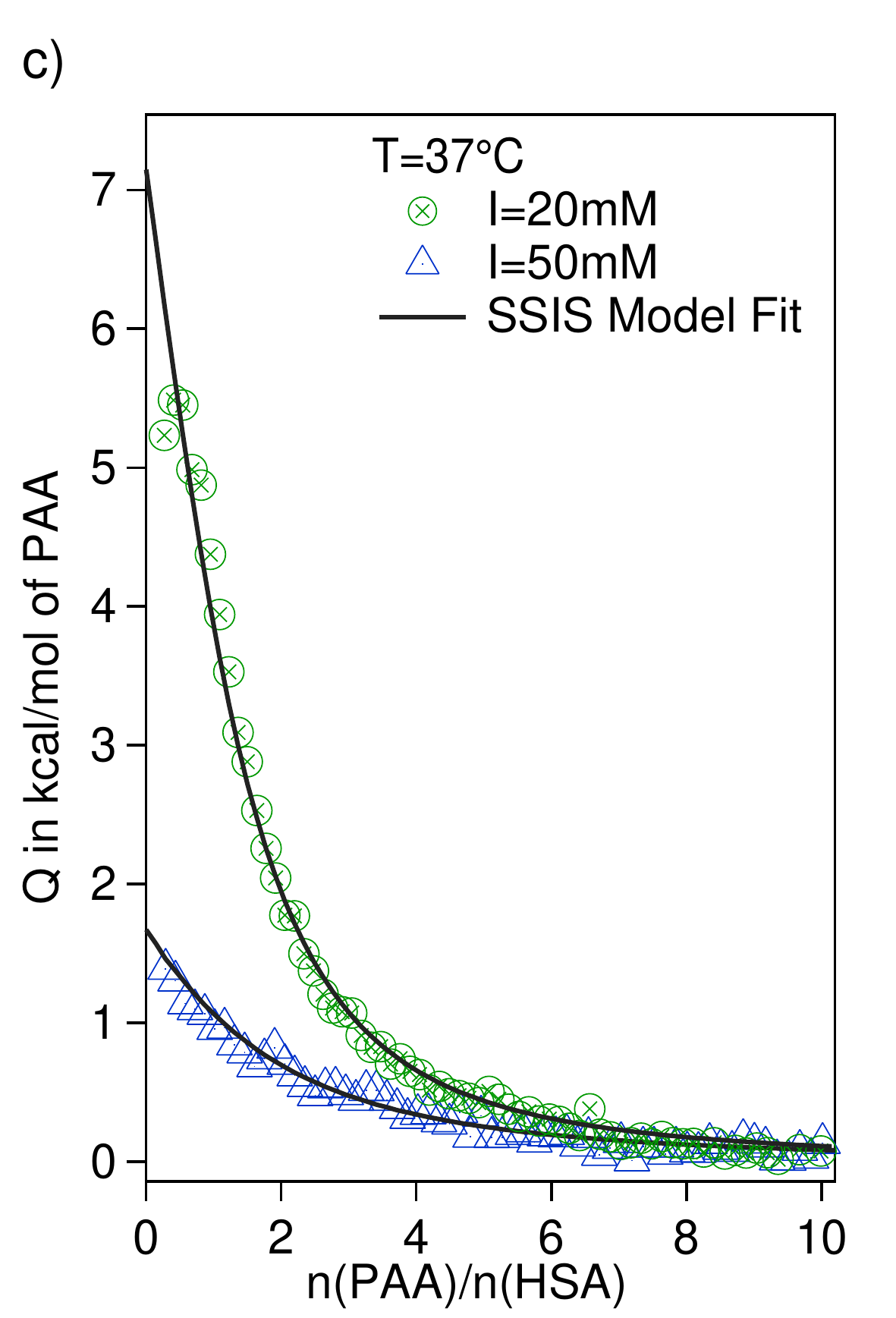}
\caption{ITC data of adsorption of PAA upon HSA and the corresponding heats of dilution of PAA at pH=7.2, T=37$^\circ$C and a) I=20~mM, b) I=50~mM. c)~Binding isotherm corrected for the heat of dilution at 37$^{\circ}$C and I=20~mM \& 50~mM.}
\label{Fig:ITCRAW}
\end{figure*}
To quantify the number of bound and released ions upon complexation, we count the average number of ions $N_i$, $i=\pm 1$, that are bound ('condensed') on the PAA chain or on the positive protein patches, respectively, and make a comparison before and after PAA/HSA association. An ion is defined as 'condensed' if it is located in the first binding layer, that is closer than a  cut-off distance $r_s = 0.5$~nm from the charged bead, while double-counting in overlapping volumes is avoided. The average is over long (ca. 30 ns) trajectories in the fully separated and stable bound states.
\begin{figure*}[!b]
\center
\includegraphics[width=0.32\textwidth]{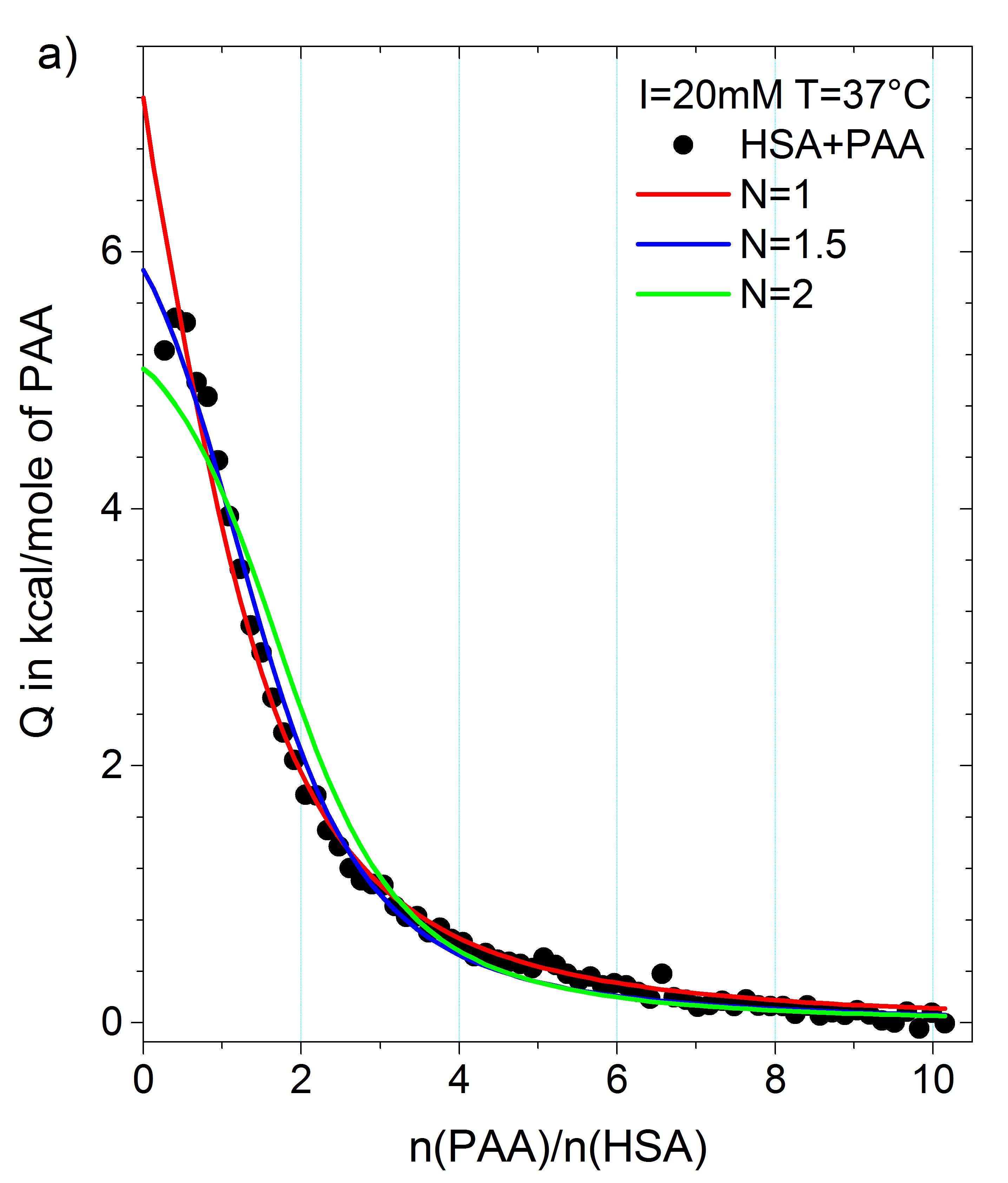}
\includegraphics[width=0.32\textwidth]{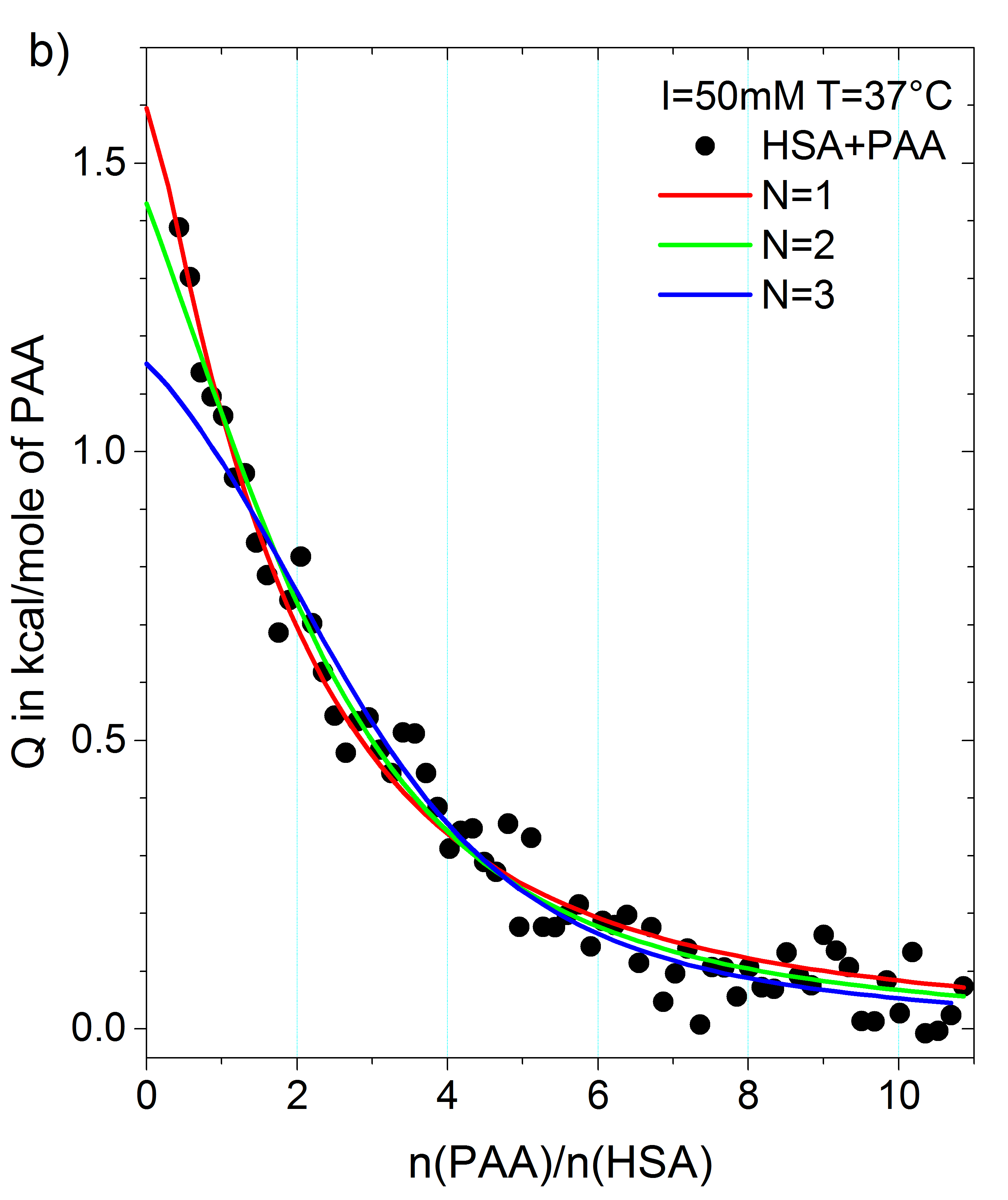}
\caption{Determination of the parameter $N$, i.e., the number of PAA molecules adsorbed on on HSA molecule is shown  for one temperature 37$^{\circ}C$ at a) I=20~mM and b) 50~mM. Each line shows a fit with different values of fixed $N$ marked by different colors.}
\label{Fig:NfitVgl}
\end{figure*}
To calculate the potential mean force (PMF) between the HSA and the PAA we  employed steered Langevin Dynamics simulations using the pull code as provided by GROMACS.~\cite{Hess2008}  Here, the center of mass of the PAA is restrained in space by an external time-dependent force. This force is applied as a constraint, \textit{i.e.} harmonic potential, and moved with a constant pulling velocity $v_p$ to steer the particle in the prescribed direction.\cite{Isralewitz2001} After several test runs, the pulling rate $v_p = 0.1$ nm/ns was chosen and a harmonic force constant $K = 2500$ kJ mol$^{-1}$ nm$^{-2}$. The simulations were performed for $\sim$100 ns. Given the pulling speed above, this simulation time is required to bring the two macromolecules from a separated state ($r \sim 11$ nm) to a final state ($r \sim 1$ nm). The standard deviation was calculated by standard block averages to specify the statistical error. The friction force {$F = -m\xi_i v_p$} was subtracted from the constraint force and the result averaged within a specific interval of discrete spacing $\Delta r$ to obtain the mean force of the interaction potential. According to the simulation setup, the mean force was integrated backwards to get the potential of mean force (PMF). Because the CPPM were radially constraint in 3D  space, the PMF has to be corrected for translational entropy~\cite{Neumann1980} by
\begin{equation}
 G(r) = G^\text{I}(r) - 2\text{k}_\text{B}\text{T}\ln(r),
\end{equation}
where $G^\text{I}(r)$ is the integrated mean force.
The binding affinity of the PAA can then be defined as the free energy value at the global minimum of the PMF in the stable complex as 
\begin{equation}
\Delta G^{sim} = G(r_{min}) - G(\infty),   
\end{equation}
where the reference $G(\infty)$ is set to zero. However, before making a comparison with the experiment, we had to consider that  $\Delta G_b^{exp}$ provided by the experiment is defined as a standard free energy, which refers to the standard binding volume $V_0 = 1/C^0$ of one liter per mol.~\cite{zhou:review} Hence, the binding constant $K_b$ generated from experiment is formulated as  $K_b = e^{- \beta \Delta G^{sim}} V_0$.  In our simulations we average the accessible volume $V_b$ of the COM of the PAA in the bound state.  

As a result, the standard binding free energy from the simulation can be obtained as~\cite{zhou:review}  
\begin{equation}
\Delta G_b^{sim} = \Delta G^{corr} + \Delta G^{sim},
\end{equation}
with a term $\Delta G^{corr} = -k_B T \ln(C^0 V_b)$ is the entropy correction arising from the accessible volume of the COM of the PAA in the bound~state. 

\section{Results and Discussion}

We performed a systematic  series of ITC experiments comprising four ionic strengths and five different temperatures ranging from room temperature (25$^{\circ}$C) to the physiological temperature (37$^{\circ}$C). The experiments were performed at pH 7.2 in buffer solution. This pH is well above the isoelectric point of the protein HSA used for this experiment leading to a net effective charge of -14. PAA is a weak acid with a pK$_a$ of 4.5 and hence a net charge of -25 at pH 7.2. Thus, we study the adsorption on the "wrong side" of the isoelectric point where both the protein as well as the polyelectrolyte is charged negatively.

ITC is certainly the method of choice for the determination of the adsorption constant of a given substrate to HSA.~\cite{Bouchemal2008,Ball2009,Kabiri2014} However, the concentrations of the protein and PAA are small and the evolved heat will be concomitantly small. Hence, all effects leading to spurious heat signals must be considered in detail and carefully excluded. The main problem is the adjustment of the same pH and ionic strength in both the solution of the protein and of the polyelectrolyte. This is done by extensive dialysis which turned out to be decisive for obtaining meaningful ITC-data. Since the concentrations of both PAA and HSA are low, all spurious effects leading to a heat signal must be considered. Since PAA-solutions are added in small portions to the solution of HSA, the heat of dilution of PAA must be determined carefully and subtracted from the raw signal.
\begin{figure*}[t]
\center
\includegraphics[width=0.3\textwidth]{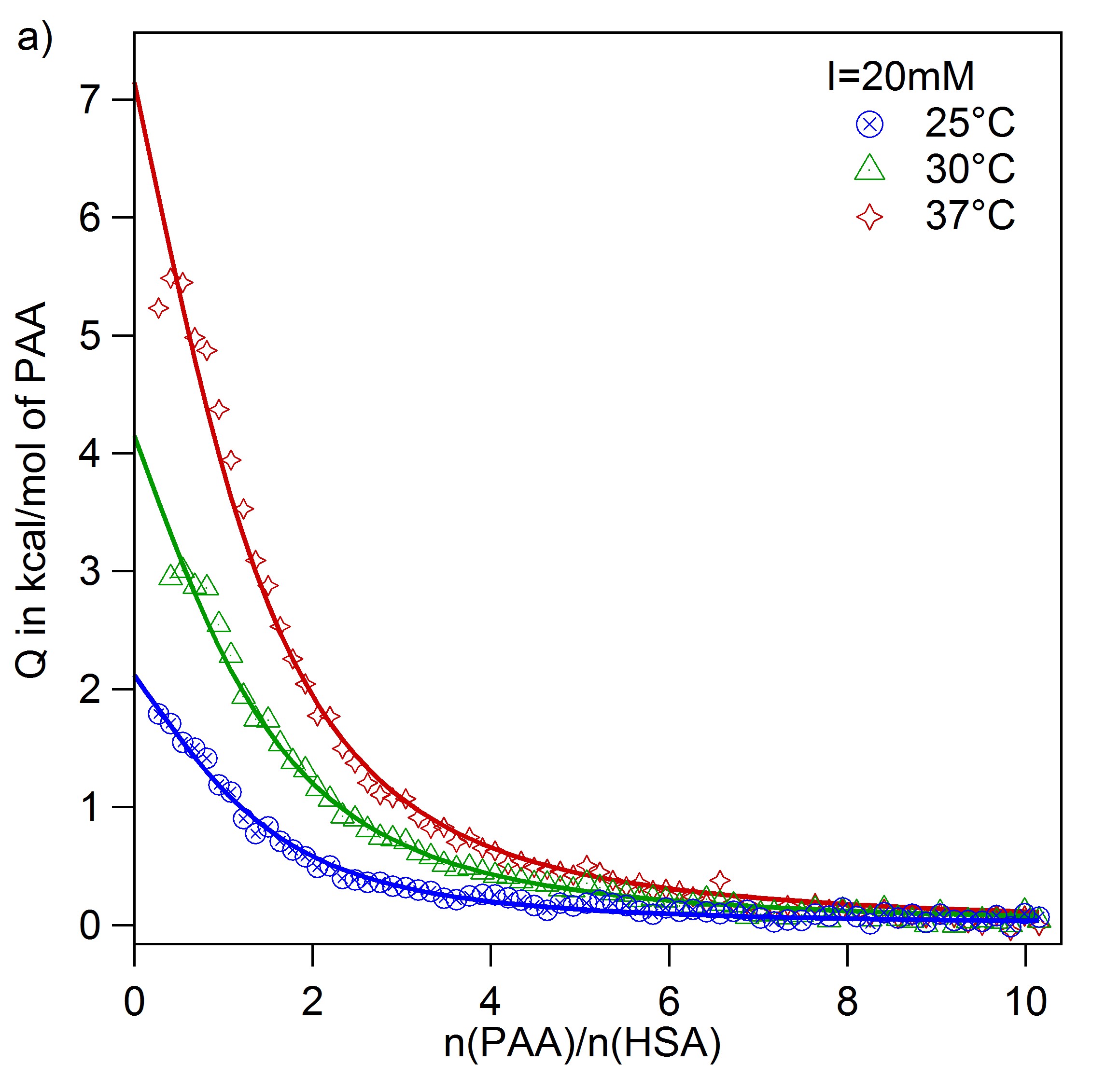}
\includegraphics[width=0.3\textwidth]{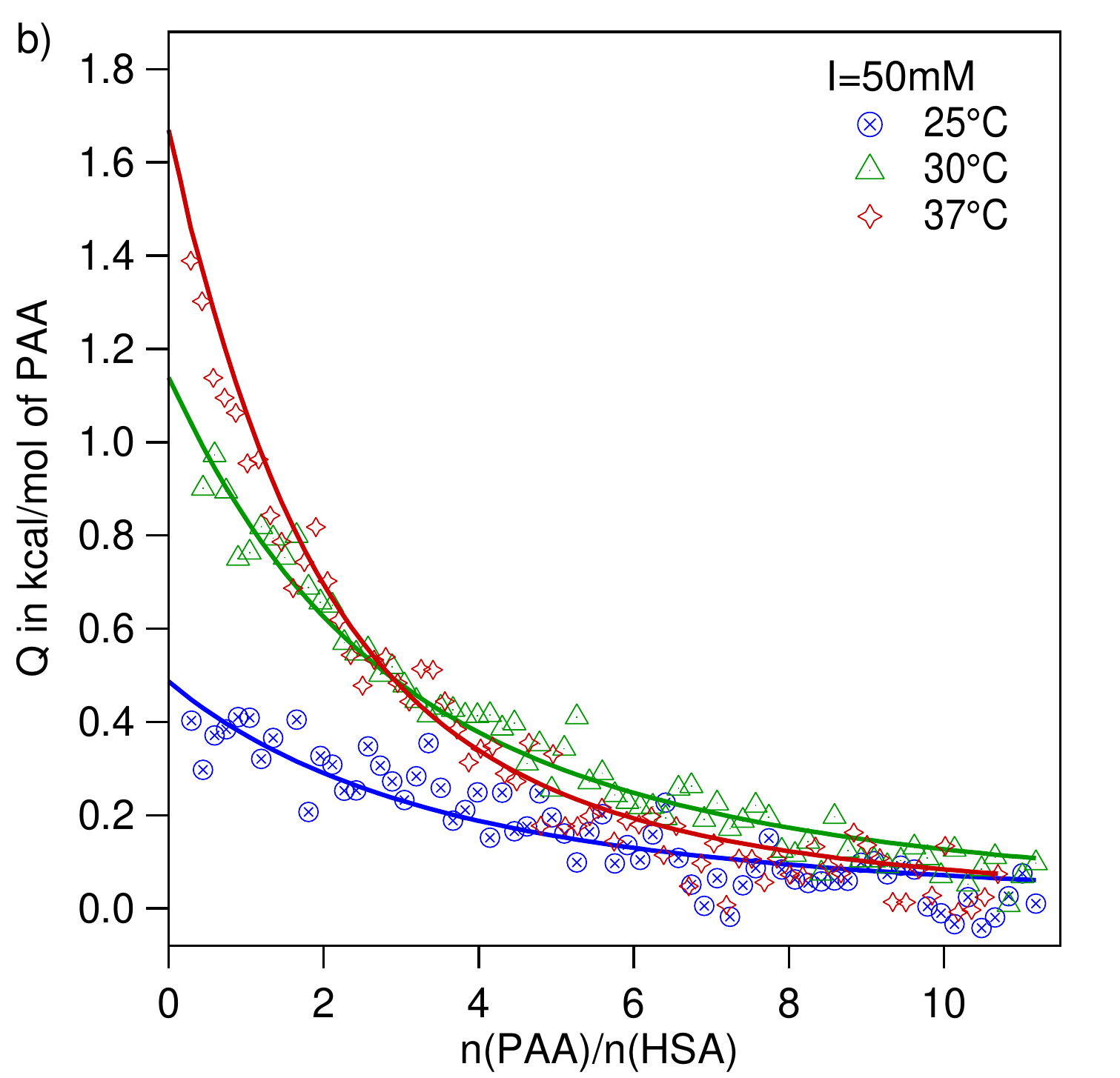}
\includegraphics[width=0.38\textwidth]{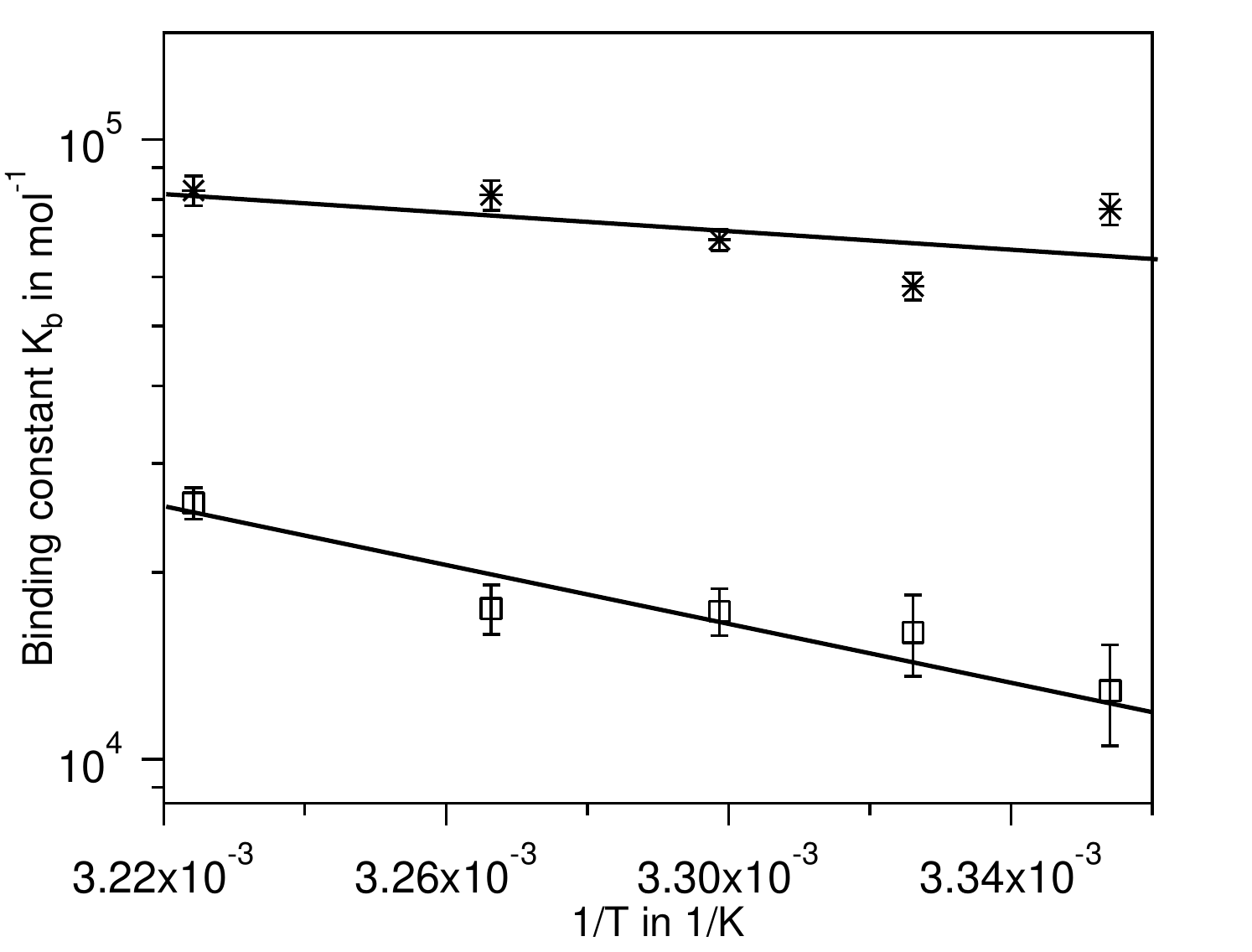}
\caption{Effect of temperature. The integrated heats Q for adsorption of PAA on HSA at temperatures between 25$^\circ$C and 37$^\circ$C for a) I=20~mM and b) I=50~mM and the according fits are displayed. For better clarity, only 3 temperatures are displayed. c) Van't Hoff analysis of the dependence of the adsorption constant on temperature. Data points are derived from the fits of the ITC-data shown in a) and b). The crosses correspond to I=20~mM and the empty squares to I=50~mM.}
\label{Fig:ITC_Qfit}
\end{figure*}
Fig.~\ref{Fig:ITCRAW} and Fig.~S1 of the SI displays the raw ITC-signals of PAA onto HSA (black curves and points) as well as the heat of dilution of PAA (green points). The signal is weakly endothermic at I=20~mM but exothermic at I=50~mM (see Fig.~\ref{Fig:ITCRAW}). Dilutions are in all cases exothermic and the effect becomes stronger with increasing salt content as expected (blue curves and points). For higher ionic strength, the heat of dilution has a dominant effect on the overall signal and determines the sign of the signal at I>20~mM. For data analysis, the heats of dilution are subtracted from the heats of adsorption prior to fitting. Here, special attention must be paid to this step, as for some cases there remains a constant residue  after subtraction of the heat of dilution. Even though this offset signal is very small and usually less than 0.1~kcal/mol, it cannot be neglected due to the small overall heat. We assign this offset to a slight mismatch of the pH or salt titrant and the solution in the cell. In order to take this effect into account, a flat background was fitted to all isotherms after the first step of subtraction and used to correct the data in the second step. 

The panel on the right-hand side of Fig.~\ref{Fig:ITCRAW} displays a set of typical results thus obtained. Evidently, the heat of adsorption is positive and we find this for all conditions under consideration here. Hence, the driving force for the process of adsorption must be entropic. This point will be discussed in more detail below and is well borne out from the simulations, too. 

In order to obtain the number $N$ of PAA-molecules bound to one HSA-molecule, the data were first fitted using the SSIS model as described in Section~2.1.1. Fig.~\ref{Fig:NfitVgl} shows a comparison of the parameter $N$ for the two data sets as in Fig.~\ref{Fig:ITCRAW}. The colored curves showing different fixed values of $N$ reveal that the data clearly justify $N=1$ as the best choice for fitting. This observation is true for any other sets of data. Deviations from $N=1$ are not significant and we can safely assume $N=1$ in all subsequent analysis (see eq.~\ref{eq:KbN1} in Section~2.1.1).
\begin{table*}
\center
\begin{tabular}{llccccc}
 \hline \\ 
    \renewcommand{\arraystretch}{1.8}
Ionic Strength  (mM) & T  ($^{\circ}$C) & $\Delta H_{ITC}$  & $K_b \cdot 10^4$ & $\Delta G_b^{exp}$ & $\Delta S_b$ & $\Delta H_b$ \tabularnewline
 & & 	(kJ/mol) &	(mol$^{-1}$) 	&	(kJ/mol) &	(kJ/mol/K) & (kJ/mol) \\
\hline \\
    \renewcommand{\arraystretch}{2.8}
   & 25	  & 16.4$\pm$0.3 & 7.7$\pm$0.5 & -27.9$\pm$1.3 &    & \tabularnewline
   & 27.5 & 32.8$\pm$0.6 & 5.8$\pm$0.3 & -27.4$\pm$1.4 &  		 	 &\tabularnewline
20 & 30   & 34.0$\pm$0.5 & 6.9$\pm$0.3 & -28.1$\pm$1.1 & 		 0.17$\pm$0.01& 15$\pm$4 \tabularnewline
   & 33   & 45.6$\pm$0.9 & 8.1$\pm$0.5 & -28.8$\pm$1.6 & 	 	 & \tabularnewline
   & 37   & 53.4$\pm$1.0 & 8.3$\pm$0.5 & -29.2$\pm$1.5 &   &\tabularnewline
   &  	  & 			 & 			   & 			   &   &\tabularnewline
&  25  & 13.6$\pm$1.1  & 1.3$\pm$0.2 & -23.4$\pm$0.5 &  &  \tabularnewline
& 27.5 & 14.4$\pm$0.9  & 1.6$\pm$0.2 & -24.2$\pm$0.3 &  &		\tabularnewline
50 &  30  & 24.8$\pm$0.7  & 1.7$\pm$0.1 & -24.6$\pm$0.1 & 0.27$\pm$0.02 & 44$\pm$8\tabularnewline
&  33  & 27.6$\pm$0.9  & 1.8$\pm$0.2 & -24.9$\pm$0.2 &  &	\tabularnewline
&  37  & 26.4$\pm$0.6 & 2.6$\pm$0.1  & -26.2$\pm$0.1 &  &   \tabularnewline
   &  	  & 			 & 			   & 			   &   &\tabularnewline
70 &  37& 21.9$\pm$0.7  & 1.0$\pm$0.1 & -23.6$\pm$0.5 & 0.12 (37$^{\circ}$C) &\tabularnewline
   &  	  & 			 & 			   & 			   &   &\tabularnewline
100 &  37 & 35.5$\pm$11.6 & 0.3$\pm$0.1 & -20.7$\pm$1.2 & 0.10 (37$^{\circ}$C)&\tabularnewline
 \hline
\end{tabular}
\caption{Overview of thermodynamic parameters for all fitted isotherms for the temperature series between 25$^{\circ}$C-37$^{\circ}$C and ionic strengths between I~=~20~mM - 100~mM. As discussed in Sec. 3, all data were fitted with fixed $N=1$. $\Delta G_b^{exp}$, $\Delta S_b$ and  $\Delta H_b$ were calculated according to eq.~\ref{eq:DeltaG},\ref{eq:Sb_vanthoff} and \ref{eq:vantHoff} respectively. Entropys for 70~mM and 100~mM were calculated using eq.~\ref{eq:Gb_1} at 37$^{\circ}$C.}
\label{Tab:Iall_N1}
\end{table*}
\subsection{Strength of interaction as the function of temperature}
\label{sec:Temperature}
Fig.~\ref{Fig:ITC_Qfit} presents two series of temperature dependency studies for the ionic strengths I=20~mM and 50~mM.
For better clarity, only 3 temperatures are displayed in these graphs. The data for two more temperatures are displayed in Figure S1 of the Supporting Information. The data taken at both ionic strength reveal a significant increase of enthalpy with increasing temperature from 25$^{\circ}$C to 37$^{\circ}$C. This effect is more pronounced for I=20~mM than at I=50~mM. Additionally, the overall enthalpy of adsorption becomes weaker with increasing salt content, which points directly to the importance of electrostatic interaction on the binding of PAA onto HSA. All data are very well described by the model assuming $N = 1$. Data taken at higher salinity are more noisy but the raise of the signal with temperature is clearly discernible. Since there is no plateau in the ITC signal, the parameter $\Delta H_{ITC}$ might be overestimated by the fits.

The results of these fits are listed in Table~\ref{Tab:Iall_N1} with $\Delta H_{ITC}$ and $K_b$ as fitting parameters, $N$ being fixed to unity (see above). The free energy of binding $\Delta G_b$ was calculated from the fitting parameter K$_b$ using equation~\ref{eq:DeltaG}. \\

The temperature dependence of the binding can now be analyzed according to van't Hoff's law: 
\begin{equation}
\left(\frac{\partial ln K_b}{\partial T^{-1}}\right)_p = -\frac{\Delta H_{b}}{R}
\label{eq:vantHoff}
\end{equation}
Fig.~\ref{Fig:ITC_Qfit}~c) shows the resulting van't Hoff plot. A linear correlation between logarithm of the binding constant $K_b$ and the inverse temperature is seen within the limits of errors. The binding enthalpy $\Delta H_{b}$ can be obtained from the slope of the linear fit and $\Delta S_{b}$ from the intercept. The resulting data are gathered in Table~\ref{Tab:Iall_N1}. In general, the values of $\Delta$H$_{ITC}$ are larger than the data resulting from the van't Hoff analysis. Similar finding have been made in a recent study of the interaction of proteins with charged microgels.~\cite{Welsch2012} Reasons may be sought in additional processes as e.g. the hydration of freed counterions that are not directly coupled to the process of binding (see below). Also, the heat of adsorption taken directly from the ITC data might be slightly overestimated (see above).


\subsection{Dependence on ionic strength.} 
\label{sec:ionicstrength}
\begin{figure}[h]
\center
\includegraphics[width=0.35\textwidth]{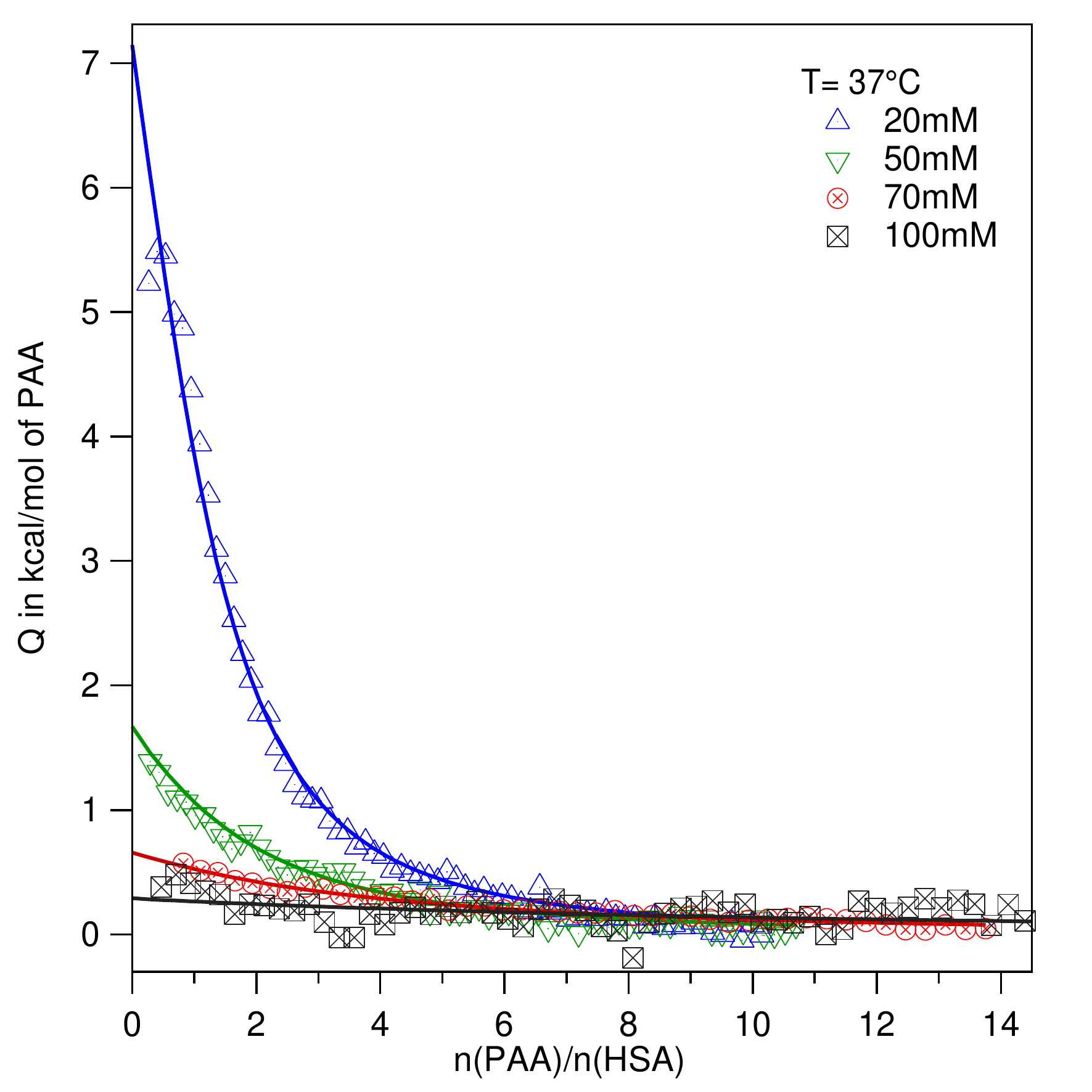}
\caption{Effect of ionic strength. Isotherms are shown for a series of ionic strengths ranging from I=20~mM-100~mM at 37$^{\circ}$C. All fits have been done with $N=1$. The thermodynamic data resulting from these fits are listed in Table~\ref{Tab:Iall_N1}.}
\label{Fig:QFit_I}
\end{figure}

To study the dependence of the binding process on ionic strength, we carried out two more experiments at 37$^{\circ}$C and I=70~mM and 100~mM (see Fig.~\ref{Fig:QFit_I}, raw data are shown in Figure S2). With increasing ionic strength, the measured enthalpy approaches zero and the ITC method reaches its instrumental limits. As described above,  we fix the parameter $N$ for both salt concentrations to unity. Table~\ref{Tab:Iall_N1} gathers all data obtained from these experiments.

\begin{figure}[h]
\center
\includegraphics[width=0.4\textwidth]{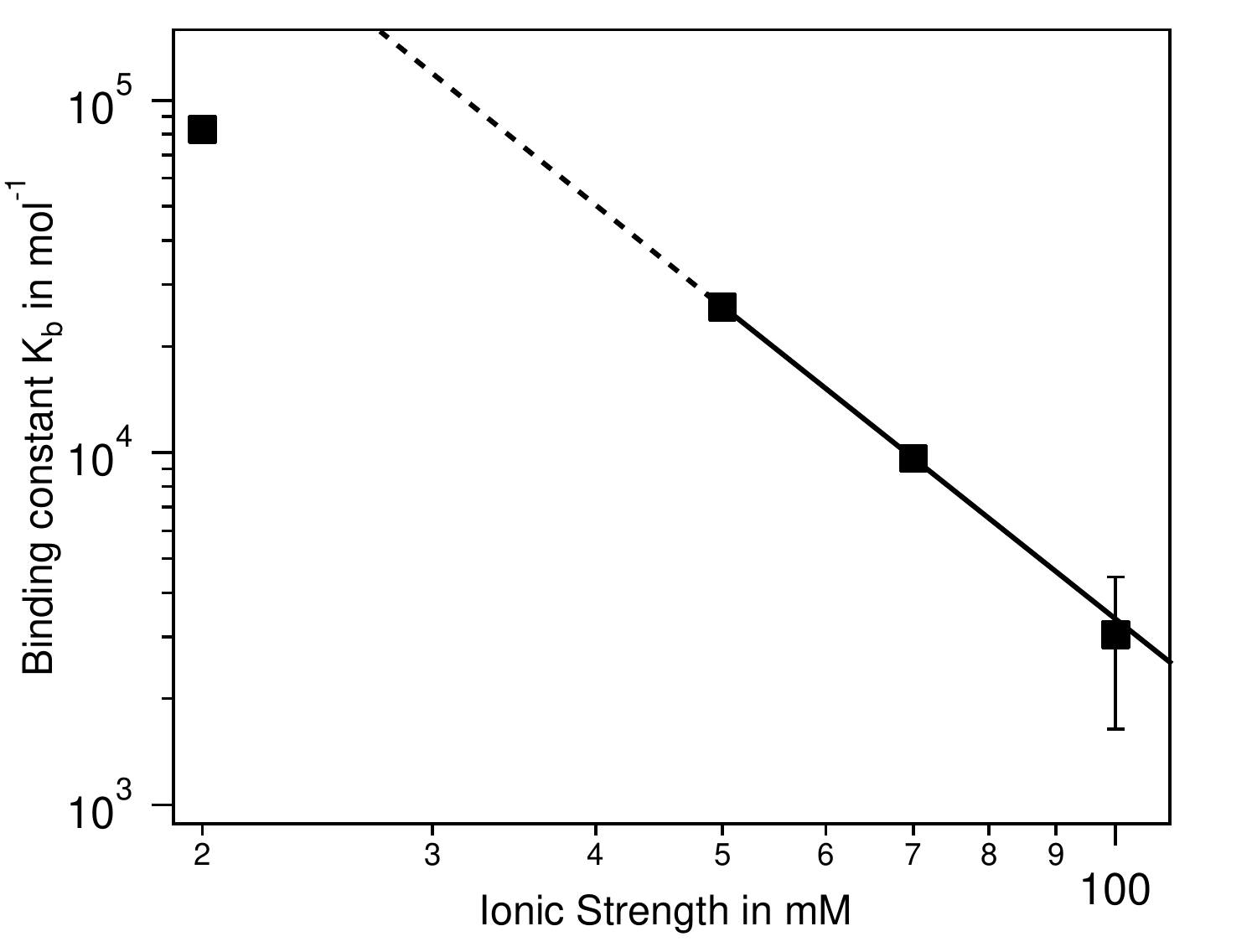}
\caption{Linear dependence of the binding constant versus ionic strength as listed in Table~\ref{Tab:Iall_N1} are depicted in logarithmic scales according to  eq.\ref{eq:Lohman}.~\cite{Record1973}}
\label{Fig:LogK_c}
\end{figure}

The data exhibit a very consistent decrease of binding constant $K_b$ with increasing salt concentration (see Tab.~\ref{Tab:Iall_N1}). This observation combined with the fact that only about one PAA molecule is adsorbed on the HSA leads to the conclusion that the driving force of the interaction is an attractive electrostatic potential between the negative PAA with patches of positive charge on the surface of HSA molecule. These patches are known to act as multivalent counterions for the polyelectrolyte. Binding of such a polyelectrolyte to such a patch thus leads to a release of its counterions.~\cite{Becker2011, Welsch2012, Henzler2010} This ``counterion release force`` was considered many years ago by Record and Lohman who predicted a linear correlation between the logarithms of binding constant and salt concentration~\cite{Record1973}: 
\begin{equation}
\frac{\text{d}ln K_b}{\text{d}ln c_{salt}}=-\Delta n_{ion}
\label{eq:Lohman}
\end{equation}
where K$_b$ is the binding constant, c$_{salt}$ the salt concentration and $n_{ion}$ the number of counterion release upon adsorption. This behavior is observed for the present data as well if the ionic strengths is above 20~mM (see Fig.~\ref{Fig:LogK_c}). Application of eq.~\ref{eq:Lohman} to the data in Fig.~\ref{Fig:LogK_c} yields $\Delta n_{ion}\approx 2.9\pm 0.5$, that is, approximately 3 ions are released upon binding of one PAA-molecule to a HSA molecule. The deviation from linearity for low ionic strength has been observed as well by Dubin~\textit{et al.} for the interaction between Bovine Serum Albumin and the polyanion Heparin at pH 6.8.~\cite{Seyrek2003} Here, electrostatic interactions become long-ranged and the relative contributions of the counterion release mechanism significantly decrease.

Extrapolation of the linear fit reveals that interaction still persists at physiological ionic strengths and temperature. At I=150~mM and 37$^{\circ}$C, we derive from our plot a finite binding free energy of around -17~kJ/mol and it only decays to small values at around 750~mM. This concentration has been found to be necessary in dialysis to remove protein-bound uremic retention solutes.~\cite{Boehringer2015} Evidently, there is still some binding under physiological conditions and much higher salt concentrations are needed to remove the toxins from the surface of HSA. These results show also that ITC-experiments can be deceptive when enthalpies of different reaction in the system compensate in such a way that they vanish. Thus, in the present case interaction still exists under physiological conditions but does not lead to measurable enthalpies. 

\section{Comparison with computer simulations}
All simulations presented in the following are using an implicit-water coarse-grained (CG) structure-based model, where the CG protein is held in its native structure by semi-empricial force-fields.~\cite{calpha}  Here the amino acids, the PAA monomers, and the salt ions are modelled explicitly on a single bead level. Water is modelled by a dielectric background continuum while the salt ions are explicitly resolved. Similar methods have been used repeatedly to study, e.g., protein folding~\cite{Lammert2009} and the pair potential~\cite{Lund2008} between proteins. They provide a reasonable picture of the interactions of a given molecule with the amino acids localized at the surface of the protein because they retain native structure, thermal motion, and ion-induced mechanisms in a well-resolved fashion. Hence, this type of simulations provides a full microscopic picture of the interaction of PAA with HSA, in particular the identification of the sites where PAA docks on. Moreover, the simulations can be used to obtain realistic free enthalpies of binding as will be further shown below. 
\begin{figure}[h]
\center
\includegraphics[width=0.35\textwidth]{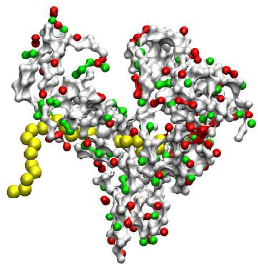}\hspace{1cm}
\includegraphics[width=0.35\textwidth]{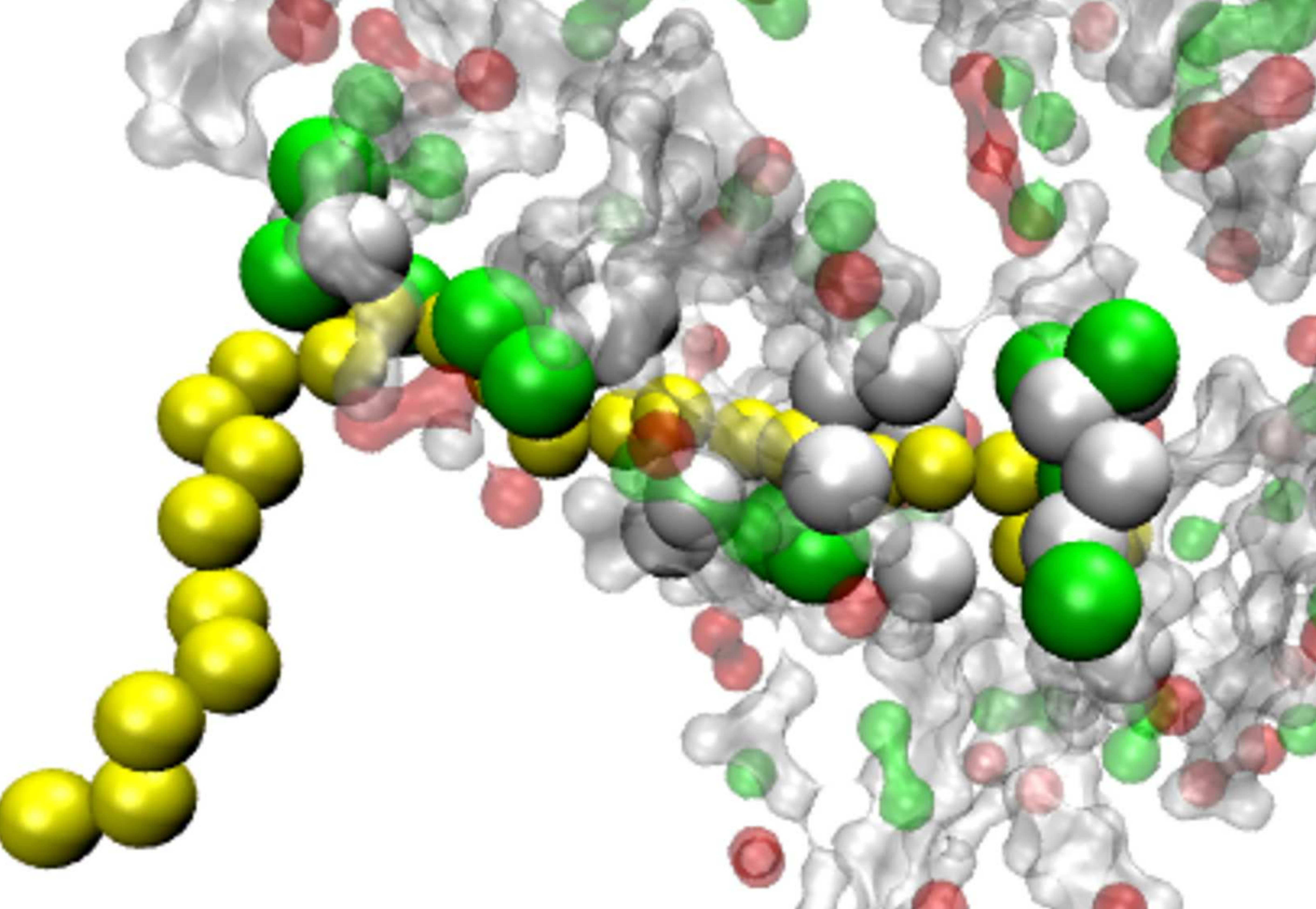}
\caption{(a) Representative computer simulation snapshot of the total HSA-PAA complex. (b) Magnification of the binding site: the PAA (yellow string of beads) is bound near the Sudlow II site. The amino acid beads that directly participate in the binding (defined by being within 0.5~nm distance to the PAA on average) are depicted by the opaque spheres. The rest of the HSA structure is distinguished by a transparent surface plot. Electrostatically neutral HSA  beads are colored white, positive beads are green, and negative beads are red. }
\label{fig:snapshot}
\end{figure}
Our computer simulations demonstrate that the HSA binds only one PAA chain, independent of temperature, salt concentration, and molar ratio in the considered parameter ranges. Hence, it reconfirms the result obtained previously by ITC. A  representative simulation snapshot of the bound complex with one PAA is presented in Fig.~\ref{fig:snapshot}. 
From a thorough screening of our simulation trajectories it emerges that this structure is highly reproducible and assumed in 80\% of the simulation time. It is hence a highly stable and probable configuration. Additional analysis reveals that the PAA chain spans the sub-domains II A, III A, and III B, involving the Sudlow II binding site. As expected for a negatively charged polyelectrolyte we find that it favorably binds positively charged amino acids, arginine (R) and lysine (K) at positions  R410, R484, R485, R413, R538, K541, K199, K195, see also the green opaque spheres in Fig.~\ref{fig:snapshot}. This is certainly a central result of the present analysis inasmuch as it defines the precise location of the binding of a highly charged molecule as PAA. 

\begin{figure}[h]
\center
\includegraphics[width=0.48\textwidth]{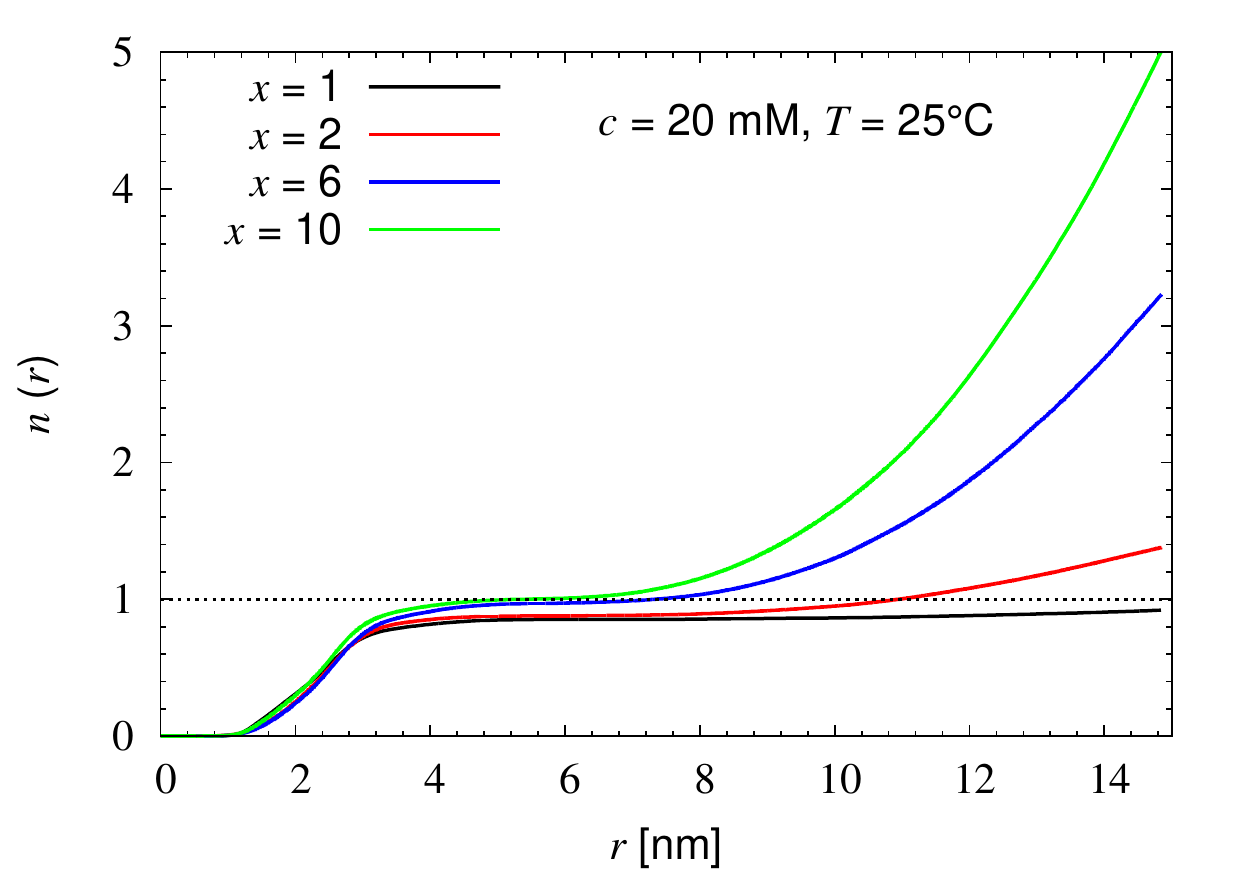}
\caption{Running coordination number $n(r)$ of the PAA chains around the HSA versus their centers-of-mass distance $r$ at a temperature of 25$^\circ$~C, salt concentration of 20~mM, and molar ratios $x=c_{\rm PAA}/c_{\rm HSA}$ = 1, 2, 6, and 10, see legend. In all cases roughly one PAA chain (horizontal black dotted line) is bound to HSA.}
\label{fig:coordination}
\end{figure}

To demonstrate that only one PAA forms a complex with HSA, the running coordination number $n(r)$ of the PAA around HSA is shown in Fig.~\ref{fig:coordination} for molar ratios $x=1$, 2, 6, and 10. The quantity $n(r)$ is the total number of PAA-molecules around a given HSA as the function of distance. The plateau of the curves between separations of $r=4$ and 6~nm just above the average value of the HSA radius is a proof that the binding number does not exceed one, irrespective of the molar ratio. Only at larger distances  $n(r)$ is increasing beyond unity since here the entire solution is explored. We find qualitatively similar results for the other investigated salt concentrations and temperatures. This finding again is in direct agreement with the results of the ITC-experiments (see also the normalized density profiles in Fig.~S5).

It is furthermore interesting to see the temporal evolution of the complex of HSA with PAA. The PAA chain slides along the Sudlow II site much in a way of a threading through an orifice. A series of snapshots combined with a movie can be found the in Fig.~S4 of the supplementary information. At the one hand, this fact demonstrates the strong binding of PAA by this side. On the other hand, the threading through this site leads to a strongly increased number of configurations of the complex and thus increases the entropy of the complex. This fact certainly leads to the binding of the PAA-chain at the Sudlow II site and not on other positive patches on the surface of HSA.

\begin{figure}[h]
\center
\includegraphics[width=0.48\textwidth]{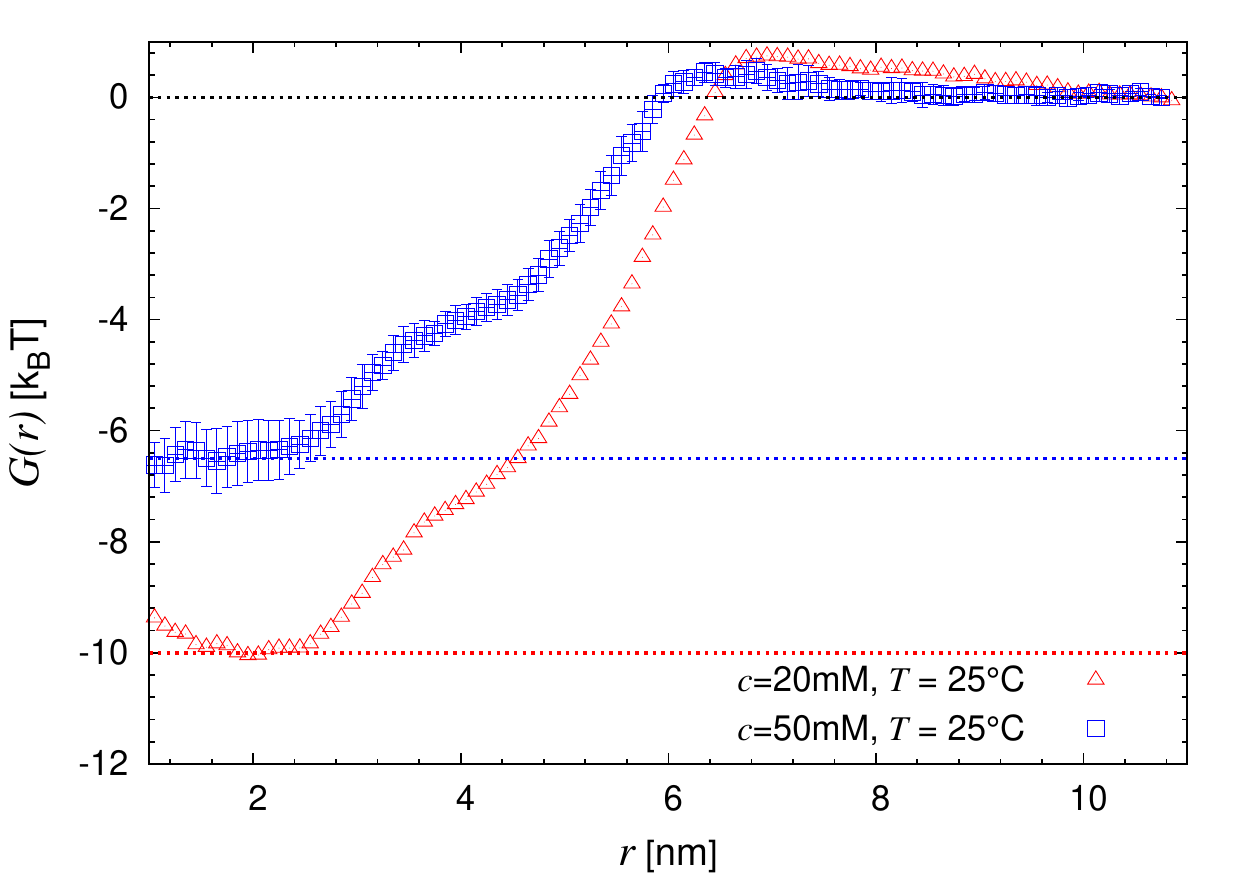}
\caption{Free energy profile (or potential of mean force) $G(r)$ between the PAA and the HSA versus their centers-of-mass distance $r$ at a temperature of 25$^\circ$~C and for 20~mM (red) and 50~mM (blue) salt concentrations. The  binding free energy $\Delta G_{sim}$ derived from the simulation can be read off as the difference between the zero free energy reference state at far separation (horizontal black dotted line) and the global minimum representing the bound state (horizontal blue and red dotted lines).}
\label{fig:pmf}
\end{figure}

Our simulations allow us to calculate the free energy profiles (potential of a mean force) along the HSA-PAA distance coordinate. Examples for this interaction free energy $G(r)$ between a single uncomplexed HSA and one PAA at two salt concentrations is presented in Fig.~\ref{fig:pmf}. For larger distances of approach, $r\simeq 7$~nm, a small repulsive barrier can be observed stemming from the monopole charge repulsion which dominates for large separations as expected. The barrier decreases and shifts slightly to shorter distances with higher salt concentration. At about $r\simeq 6$~nm the onset of a strong attraction takes place until a global minimum is observed at closer approach at about $r=2$~nm. The onset of attraction happens right at values comparable to half of the contour length of the PAA chain at which one of its ends  is first able to get in contact with the HSA surface. We never found a stable free energy minimum for the adsorption of a second PAA. This is due to a too strong monopole charge repulsion and the covering of the high-potential binding spot by the firstly bound PAA. 

For the stable HSA/PAA complex, the binding free energy $\Delta G^{sim}$ can be calculated  from the difference of the global minimum and the reference free energy  at large distances (horizontal lines in Fig.~\ref{fig:pmf}).  The values of the simulation binding free energy,  corrected to yield the {\it standard} free energy of binding from the  simulations $\Delta G_b^{sim}$ (see Methods), are summarized in  Table~\ref{Tab:FreeEnergy_theo} for various salt concentrations and temperatures. We find a very good agreement for  all systems with largest deviation of only 13\% for the case of 50~mM salt at 37$^\circ$~C temperature. As in the experiments, the binding affinity decreases in the simulations with higher salt concentrations and increases  with rising temperature.  The highly quantitative description  by the simulations, however,  is actually somewhat surprising given the simplicity of the underlying model and the neglect of hydration effects and should be discussed with caution. However, we take it as a strong indication that relatively generic electrostatic interactions rule the complexation process and hydration contributions (such as hydrophobic or van der Waals (vdW) attractions) are rather small. 

\begin{table*}
\begin{tabular}{cccccc}
\hline 
Conditions & $\Delta G^{sim}$ (kJ/mol) & $V_b$ ($nm^3$) & $\Delta G^{corr}$ (kJ/mol)& $\Delta G_b^{sim}$ (kJ/mol) & $\Delta G_b^{exp}$ (kJ/mol) \tabularnewline
\hline 
20~mM, 25$^{\circ}$C& $-24.8\pm 4.0$ & 3.0  & $-1.5$  & $-26.3\pm 4.0$ 	& $-27.9\pm 0.2$\tabularnewline
20~mM, 37$^{\circ}$C& $-25.1\pm 3.6$ & 5.9  & $-3.1$  & $-28.2\pm 3.6$ 	& $-29.2\pm 0.2$\tabularnewline
50~mM, 25$^{\circ}$C& $-16.1\pm 1.0$ & 16.0 & $-5.7$  & $-21.8\pm 1.0$ 	& $-23.5\pm 0.5$\tabularnewline
50~mM, 37$^{\circ}$C& $-18.9\pm 0.5$ & 7.5  & $-3.9$  & $-22.8\pm0.5$	& $-26.2\pm 0.2$\tabularnewline
\hline 
\end{tabular}
\caption{The calculated standard binding free energy from the simulations $\Delta G_b^{sim}$ in comparison with the experimental ones $\Delta G_b^{exp}$ for various salt concentrations and temperatures in units of kJ/mol. $\Delta G^{sim}$ is the direct output from the simulations which has to be corrected by $\Delta G^{corr}$ for the binding volume $V_b$ to obtain the standard free energy of binding (see Methods)}
\label{Tab:FreeEnergy_theo}
\end{table*}

In order to test our hypothesis introduced above that the binding free energy is essentially dominated by the entropy of released counterions from the PAA chain and/or a positive patches on the HSA, we have counted the number of released ions upon complexation. In brief, we define an ion as 'condensed' if it is located in the first bound layer near a charged HSA or PAA monomer, defined by a cut-off radius of 0.5~nm. 
Hence, the number of released ions is calculated as the difference of the average number of condensed ions in the fully separated and the stable bound states (see Fig.~S5). 
 For a temperature of 25$^\circ$C and 20~mM and 50~mM salt concentrations, our analysis shows that on average indeed 2.5 condensed ions are diluted away into the bulk upon complexation. This is indeed in good agreement with the Record-Lohman-analysis of the experiment data discussed above (cf. the discussion of Fig.~\ref{Fig:LogK_c}).  

Deeper inspection shows that 2 of those ions come from the PAA, at which they where bound in a high density state (see Fig.~S6 for ionic profiles around PAA monomers). If we now just consider the PAA-condensed ions and their average concentration in the bound state $c_{dense} \simeq 1.5 \pm 0.5$~mol/l,  this implies that a favorable entropy contributions of about $\simeq k_BT \ln (c_{dense}/c_s) = 4.3\pm 0.4$ and $3.4 \pm 0.4~k_BT$ is gained per ion upon its release into 20~mM and 50~mM bulk concentrations, respectively. The total release free energies estimated by this analysis are thus roughly -21$\pm 2$ and -17$\pm 2$~kJ/mol for 20~mM and 50~mM salt, respectively, which is close  to the binding free energies from both experimental data and from simulations. Hence, the  binding of PAA is to a great part ruled by a counterion release mechanism and entropy. We note, however, that the matching of these numbers may be fortunate since other non-negligible interactions such as (repulsive) chain entropy, vdW attractions, and multipolar charge interactions beyond the bound ion layer (that is, from screening ions), all present in both simulation and experiment,  have been neglected in this simple counterion release concept. The present comparison with experimental data, however, indicates that these contributions are of  comparable magnitude and cancel each other roughly for the present system.  

Evidently, we obtained the leading contribution of the ions directly condensed on the PAA chain. Hence, the estimate of the entropy of  counterion release given above should be considered a lower bound for the absolute entropy  contribution.  Other contributions not included in the theoretical analysis apparently lead to experimental entropies that are higher by a factor of 2-3 (see cf.~Table~\ref{Tab:Iall_N1}). The good agreement between the measured and the calculated  $\Delta G_b$, however, demonstrates that these additional entropic contributions are canceled out by an enthalpic contribution of equal magnitude. This ``enthalpy-entropy compensation`` is well-known for various processes as e.g. solute hydration, protein folding, or proteins association.~\cite{Ball2002,Kabiri2014} The present comparison of theory and experiment allows us to discern among these terms and the leading contribution to $\Delta G_b$.

\section{Conclusions}
We presented a study of the binding of short PAA chains to HSA by combining calorimetry, that is, ITC and computer simulation. Both ITC experiments and simulation results show that there exists a strong attractive interaction between PAA and HSA at ``the wrong side of pI``, where both are negatively charged. Computer simulations demonstrate that the binding of the PAA takes place at the Sudlow II site. ITC measurements for a series of salt concentrations between 20~mM and 100~mM show that the dependence of binding affinity $\Delta G_b$  on ionic strengths is mainly determined by the counterion release mechanism and can be described by the Record-Lohman relation see eq.~\ref{eq:Lohman} (see Fig.~\ref{Fig:LogK_c}). The binding affinity $\Delta G_b$ decreases with high ionic strength, until it practically vanishes at around 750~mM. Both the analysis of the experimental data as well as the simulations find that approximately 3 ions are released in the binding process. The binding affinity $\Delta G_b$ can be calculated from simulations with good accuracy. Combining simulations with calorimetry is hence a powerful tool to elucidate the interaction of proteins with given substrates or possible toxins. Thus, this combination of theory and experiment is now capable of solving biochemical problems of direct medical importance.

\acknowledgement
The Helmholtz Virtual Institute for Multifunctional Biomaterials for Medicine are gratefully acknowledged for financial support. The authors thank the Helmholtz Association for funding of this work through Helmholtz-Portfolio Topic "Technology and Medicine". X. Xu is sponsored by the China Scholarship Council (CSC). In addition, J. Jankowski was supported by the grant "NPORE" of the BMBF/IGSTC (01DQ13006A).


\bibliography{Bibliography}
\bibliographystyle{achemso}
\end{document}